\documentstyle[aps,epsfig]{revtex} 
\begin{document}
\draft

\title{Directed polymers at finite temperatures in 1+1 and 2+1 dimensions}

\author{Xiao-Hong Wang$^{1,2}$, Shlomo Havlin$^{1}$\footnote{Corresponding author}
 and Moshe Schwartz$^{3}$}
\address{$^1$ Minerva Center and Department of Physics,
         Bar--Ilan University,
         Ramat Gan 52900, Israel}
\address{$^2$ Department of Thermal Science and Energy Engineering,
          University of Science and Technology of China,
          Hefei, Anhui 230026, P. R. China}
\address{$^3$ School of Physics and Astronomy, Raymond and Beversly
         Sackler Faculty of Exact Science, Tel-Aviv University, Tel-Aviv
         University, Tel-Aviv 69978, Israel}

\date{\today}

\maketitle
\begin{abstract}

    We present systematic numerical simulations for directed polymers at
finite temperatures in 1+1 and 2+1 dimensions. The transverse fluctuations
and free energy fluctuations tend to the strong coupling limit at any
temperature in both 1+1 and 2+1 dimensions for long time $ t $. Two
different definitions for energy fluctuations at finite temperatures,
which are the ensemble energy fluctuations and the internal energy
fluctuations,
are investigated. Apart from zero temperature, the behavior of the energy
fluctuations and the free energy fluctuations for directed polymers is
shown
to be different. At finite temperatures, the ensemble energy fluctuations
in both 1+1 and 2+1 dimensions and internal energy fluctuations in 1+1
dimensions scale as $ t^{1/2} $ where the free energy fluctuations in
1+1 dimensions and 2+1 dimensions scale as
$ t^{1/3} $ and $ t^{0.2} $ respectively.
As a consequence of that the specific heat in both 1+1 and 2+1
dimensions scales as $ t $  and the entropy fluctuations in 1+1
dimensions scale as $ t^{1/2} $ at any finite temperature.

\end{abstract}



\vskip 0.1in
\noindent {\bf Introduction}
\vskip 0.1in
    Rough surfaces and interfaces are found in a wide variety of natural
and industrial processes. The KPZ equation provides a good quantitative
understanding of surface growth phenomena$ ^{1- 5} $. It is well known
that the KPZ equation is equivalent to the problem of directed polymers in
a random medium$ ^{6- 9} $, which has attracted much attention in recent
years. The problem of directed polymers is also relevant to many other 
fields ranging from paper rupture$ ^{10} $ and spin glasses$ ^{11} $ to
high-$ T_{c} $ superconductors$ ^{12} $. The KPZ equation provides the
exact dynamical exponent for directed polymers in 1+1 dimesions$ ^{6- 8} $.
A self-consistent expansion of the correlation function$ ^{13} $ yields
results compatible with simulations for 2+1 dimensions. Above 2+1
dimensions, the behavior of directed polymers has a phase transition when
temperature is raised and the system goes over from the strong coupling
behavior to a perturbative behavior$ ^{13- 18}$. The strong coupling
behavior has been studied by extensive numerical investigations for the
energy fluctuation at zero temperature$ ^{6,8, 19-21} $, which is ensemble
fluctuations of the ground state energy of directed polymers.
At finite temperatures, fluctuations of the free energy have been studied
extensively but it seems that quantities like fluctuations of the internal
energy have not received much attentions.

    In this article, we present systematic numerical simulations for
directed polymers in 1+1 and 2+1 dimensions.
Our numerical simulations show that the transverse fluctuation
and free energy fluctuation crossover to the strong coupling behavior at
any finite temperature in both 1+1 and 2+1 dimensions, in agreement with
the predictions of the field theory$ ^{13,22,23} $. 
We define the energy fluctuations at finite temperatures using two
different definitions: the ensemble energy fluctuations and the internal
energy fluctuations. The difference between the two definitions is related
to the specific heat. We find that the behavior of the energy fluctuation
at finite temperatures is different from that at zero temperature. For
understanding the energy fluctuations at finite temperatures, we introduce
the entropy and study its fluctuation. The behavior of the energy
fluctuations at finite temperatures can be understood after investigating
the specific heat and the entropy fluctuations.

\vskip 0.1in
\noindent {\bf Model}
\vskip 0.1in

    Consider a directed polymer on a hyperpyramid lattice structure with
the random energy assigned on each bond. The partition function $ G(x,t) $
for directed polymers starting from $ (0,0) $ and ending at $ (x,t) $ is
defined by $ G(x,t)= \sum\limits_{C}e^{-E_{C}/T} $ where $ E_{C} $ is the
sum of the energies on the path $ C $ and $ T $ is the temperature. 
The iteration relation for the partition function $ G(x,t) $ is 
\begin{equation}
G(x,t+1)=G(x-1,t)e^{-\epsilon_{l}/T}+G(x+1,t)e^{-\epsilon_{r}/T},
\end{equation}
\noindent in which, $ {\epsilon}_{l} $ and $ {\epsilon}_{r} $ are the
energy assigned to the left and right bonds of the point $ (x,t). $ The
free energy $ F(t) $ is given by $ F(t)=-TlnG(t), $ where
$ G(t)=\sum\limits_{x}G(x,t) $ is the total partition function. 
The transverse fluctuation $ \Delta x $ for the directed polymers is $
(\overline{\langle x^{2} \rangle})^{1/2} $ where  
$ \langle x^{2} \rangle= {\sum\limits_{x}x^{2}G(x,t)/G(t)} $ ($
\overline{A} $ is the ensemble average of the quantity $ A $).  
The transverse fluctuation 
$ \Delta x $ and free energy fluctuation $ \Delta
F=(\overline{F^{2}}-\overline{F}^{2})^{1/2} $
have been commonly studied. 
Another interesting quantity for directed polymers is
the energy fluctuation. It is clear that the energy fluctuation at zero
temperature is just that of the free energy. For finite temperatures, we
propose two definitions for the energy fluctuations. One is the {\sl
ensemble energy
fluctuation} and the other is {\sl internal energy fluctuation}. The
ensemble energy fluctuation is defined by $ (\Delta
E)_{B} \equiv (\overline{\langle E^{2} \rangle}-\overline{\langle E
\rangle}^{2})^{1/2} $, in which $ \langle E \rangle $ is the internal
energy:
\begin{equation} 
\langle E \rangle = {\frac {\sum\limits_{x}\sum\limits_{C}
E_{C}e^{-E_{C}/T}}{\sum\limits_{x}\sum\limits_{C}e^{-E_{C}/T}}}
=T^{2}{\frac {\partial lnG}{\partial T}}
\end{equation}
\noindent and 
\begin{equation}
\langle E^{2} \rangle ={\frac
{\sum\limits_{x}\sum\limits_{C}E_{C}^{2}e^{-E_{C}/T}}{\sum\limits_{x}
\sum\limits_{C}e^{-E_{C}/T}}} = T^{2}{\frac {\partial}{\partial T}}[T^{2}
{\frac {\partial lnG}{\partial T}}]+[T^{2}{\frac {\partial lnG}{\partial
T}}]^{2}.
\end{equation}
\noindent The internal energy fluctuation is defined by $ (\Delta
E)_{T} \equiv (\overline{{\langle E \rangle}^{2}}-\overline{\langle E
\rangle}^{2})^{1/2}. $ It is obvious that both the two 
energy fluctuations will degenerate into the fluctuation of the sum of 
the energy on the optimal path when temperature tends to zero. From
the
equations (2)
and (3), we have
\begin{equation} 
(\Delta E)_{B}^{2} =(\Delta E)_{T}^{2}+T^{2}C(T,t) 
\end{equation} 
where $ C(T,t) = {\frac {\partial}{\partial T}}[T^{2}{\frac
{\partial \overline{lnG}}{\partial T}}] = {\frac {\partial 
\overline{\langle E \rangle}}{\partial T}} $ is the
specific heat$ ^{24} $. 

    In order to understand the internal energy fluctuation , we use the
entropy as $ S=(\langle E \rangle -F)/T $ and the entropy fluctuation is $
\Delta S = (\overline{S^{2}}-\overline{S}^{2})^{1/2} $. It is easy to see
that
\begin{equation}
(\Delta S)^{2}={1\over T^{2}}[(\Delta E)_{T}^{2}+(\Delta
F)^{2}]-{2\over T^{2}}B(T,t),
\end{equation}
where $ B(T,t)=\overline{\langle E
\rangle F}-\overline{\langle E \rangle}~{\overline{F}}=-{1\over 2}T^{3}
{\frac {\partial}{\partial T}}[(\Delta F)^{2}/T^{2}]. $ 

    The above quantities can be calculated by the iteration relations
given in Appendix. In the following, we present systematic numerical
simulations based upon the equation (1) and equation (A1) and (A2) from
the Appendix with the initial conditions
$ G(x,0)={\delta}_{x,0},  \widehat{E}(x,0)=0 $ and $
\widehat{E}_{2}(x,0)=0. $ The random energy assigned on the bond is
assumed to be uniformly distributed in the interval (-0.5,0.5) and
uncorrelated in space and time. The effective program RAN2 was
used for generating the uniform distributed random number$ ^{25} $. We use
length up to $ t=2000 $ $ (d=1+1) $ and $ t=150 $
$ (d=2+1) $. Six thousands configurations for 1+1 dimensions and four
thousands configurations for 2+1 dimensions were collected for the
ensemble average. In order to avoid unstability, we normalize the
partition function at every time step.

\vskip0.1in
\noindent {\bf Directed polymers in 1+1 and 2+1 dimensions}
\vskip 0.1in

At infinite temperature, directed polymers become random walks: the
transverse fluctuation $ \Delta x $
will scales as  $ t^{1/2} $  and the ensemble energy fluctuation will be $
(\Delta E)_{B}  \propto t^{1/2} $ because the ensemble energy fluctuation
at $ N $ time steps is only the variance of the sum of $ N $ uncorrelated
random numbers at the limit. The free energy fluctuation for infinite
temperature can be predicted by the Edwards-Wilkinson equation$ ^{26- 28}
$ that describes the linear surface growth phenomena. It follows that the
free energy fluctuation $ \Delta F $ scales as $ t^{1/4} $ in 1+1
dimensions and tends to the logarithmic behavior in 2+1 dimensions. 

    When approaching zero temperature, the system will take the strong
coupling behavior. Our simulations show that the transverse fluctuation
scales as $ t^{\nu} $ and both the free energy and energy fluctuations
scales as $ t^{\omega} $, in which $ \nu=2/3, \omega=1/3 $ for $ d=1+1 $
and $ \nu \approx 0.6, \omega \approx 0.2 $ for $ d=2+1 $. These results
are in agreement with the theoretical treatments and the previous
numerical works$ ^{6-9,13,19-20} $. 

    The problem of directed polymers in 1+1 and 2+1 dimensions at
finite temperatures is investigated. Our simulations show that the
transverse fluctuation $ \Delta x $ crossovers from $ t^{1/2} $ to $
t^{2/3} $ and the free energy fluctuation $ \Delta F $ crossovers from $
t^{1/4} $ to $ t^{1/3} $ in 1+1 dimensions at any finite temperature (see
Fig. 1) in agreement with earlier arguments$ ^{29} $. 
The scaling relations $ \Delta x = T^{2}f_{1}(T^{-4}t) $ and $ \Delta F =
Tg_{1}(T^{-4}t), $ which describe the crossover from the Edwards-Wilkinson 
behavior to KPZ behavior in 1+1 dimensions given by Amar and Family using
the scaling analysis$ ^{29} $ , are confirmed for $ T > 0.25 $ 
(see Fig. 2).

    Similar to 1+1 dimensions, the transverse fluctuation $ \Delta x $ and
free energy fluctuation $ \Delta F $ crossover from $ t^{1/2} $ to $ 
t^{0.6} $ and from the logarithmic behavior to $ t^{0.2} $ respectively in
2+1 dimensions (see Fig. 3). For 2+1 dimensions, we can fit the data very
well using $ (\Delta F)^{2}=0.0068ln(1+20t) $ at very high temperature.
Assuming the scaling relations $ \Delta x = 
Tf_{2}(T^{-2}t) $ for $ T >0.1 $ and $ \Delta F =Tg_{2}(T^{-2}ln(1+20t))
$ for $ T> 0.2 $, the data collapse very well (see Fig. 4). We note
that an argument for the crossover time $ lnt_{\times} \sim T^{2} $
for the free energy fluctuation has been given earlier$ ^{16, 22-23} $.
Also note that our numerical data suggest that the crossover time for $
\Delta x $ and $ \Delta F $ are different. 

    The difference between the two different definitions for the energy
fluctuation is proportional to the specific heat as shown in equation (4).
Our simulations shows that that specific heat $ C(T,t) \rightarrow 0 $
for $ T \rightarrow 0 $ as $ O(T^{\alpha}) $ where $ \alpha=1.3 $ for $
d=1+1 $ ,$ \alpha=1.6 $ for $ d=2+1 $ and $ C(T,t) \rightarrow 
{\sigma}^{2}/T^{2} $ for $ T \rightarrow \infty $  where $ \sigma=({1\over
12})^{1/2} $ is the variance of the random numbers. We also find that
the specific heat scales as $ t $ at any temperature in both
1+1 and 2+1 dimensions (see Fig. 5). Fig. 6 shows that the ensemble energy
fluctuations $ (\Delta E)_{B}^{2} $ scale as $ t $ as the specific heat
apart from zero temperature. We find very good fit with the scaling
relations $ (\Delta E)_{B}^{2}=e^{1/T}h_{1}(e^{-1.5/T}t)
$ in 1+1 dimensions and $ (\Delta E)_{B}^{2}= e^{0.2/T}h_{2}(e^{-0.5/T}t)
$ in 2+1 dimensions for $ T \leq {1\over 3} $ (see Fig. 7). Howevever, in
both 1+1 and 2+1 dimensions, the picture for the internal energy
fluctuation is not clear (see Fig. 8).  We note that the internal energy
fluctuations is not monotonic with the temperature. But, 
at the two limits of zero and infinite temperatures, the internal energy
and free energy fluctuations are the same in both 1+1 and 2+1
dimensions. For investigating the difference between the internal energy
fluctuations and the free energy fluctuations, we study the entropy
fluctuations.

    Entropy is an important physical quantity for characterizing the
disorderd systems. Fig. 9 shows that the entropy fluctuation has the
behavior  $ (\Delta S)^{2} \propto t $ at any temperature in 1+1
dimensions.  This result is in agreement with earlier arguments of
Fisher and Huse$ ^{30} $. The entropy fluctuation $ (\Delta S)^{2} $
reaches a maximum about $ T=T_{P}=0.2 $ and 
tends to zero as $ O(T^{-4}) $ when $ T \rightarrow \infty $ and as $ O(T)
$ when $ T \rightarrow 0 $ (see Fig. 10). It follows that the internal
energy fluctuation and the free energy fluctuation are the same at
infinite temperature and the energy fluctuation and the free energy
fluctuation are the same at zero temperature. 
The function $ B(T,t) $ in the equation (5) can also be calculated
numerically. From the equation (5), it is easy to see that the function $
B(T,t) $ is the same as the free energy fluctuation or the internal energy
fluctuation at the two  limits of zero and infinite temperature. At finite
temperatures, Fig. 11a shows
that the behavior of $ B(T,t) $ is similar to that of the free energy
fluctuation $ (\Delta F)^{2} $ and also has a crossover from $ t^{1/2} $ to $
t^{2/3} $ in 1+1 dimensions. 
From the equation (5), we can expect that the internal energy fluctuations
$ (\Delta E)_{T} $ will scale as to $  t^{1/2} $ as the behavior of the
entropy fluctuations $ \Delta S $ at finite temperatures in 1+1
dimensions.

\vskip 0.1in 
\noindent {\bf Discussion}
\vskip 0.1in

Using the iteration relations (1), (A1) and (A2), we study the problem
of directed polymers in 1+1 and 2+1 dimensions. The transverse
fluctuations and free energy fluctuations have a crossover from the EW
behavior to KPZ behavior at finite temperatures in both 1+1 and 2+1
dimensions as predicted by the field theory. In order to understand the
energy fluctuations at finite temperatures, we investigate the specific
heat and the entropy fluctuations. The specific heat in both 1+1 and 2+1
dimensions and the entropy fluctuations $ (\Delta S)^{2} $ in 1+1
dimensions scale as $ t $ at any finite temperature. As a result, the
ensemble energy fluctuations in 1+1 and 2+1 dimensions and the internal
energy fluctuations in 1+1 dimensions scale as $ t^{1/2} $ at finite
temperatures. It means that the energy fluctuations and the free energy
fluctuations may have a different behavior at finite temperatures in both
1+1 and 2+1 dimensions. The behavior of the entropy fluctuations in 2+1
dimensions is more complicated than that in 1+1 dimensions. There is an
indication of a phase transition in 2+1 dimensions. The phase transition
occurs at the temperature where the entropy fluctuation reaches the
maximum. This result will be published elsewhere$ ^{31} $. Fig. 11b shows
that the function $ B(T,t) $ in 2+1 dimensions also has a crossover from
the logarithmic behavior to $ t^{0.2} $ at any finite temperature. As a
result, the internal energy fluctuations in 2+1 dimensions also have a
phase transition even if Fig. 8b can not show it clearly.

\vskip0.1in

{\bf Acknowledgement.} The authors are grateful to Nehemia Schwartz and Ehud Perlsmann for useful
discussions. X.H.Wang acknowledges financial support from the Kort
Postdoctoral Program of Bar-Ilan University.

\vskip 0.4in
\noindent {\bf Appendix: Iteration relations}
\vskip 0.2in

    As in deriving the iteration relation for the partition function $
G(x,t) $ given by Eq. (1), we concentrate on the 1+1 dimensional case. The
generlization to higher dimensions is obvious. In order to obtain the
iteration relations for calculating the energy and entropy fluctuations,
we define $ \widehat{E}(x,t) \equiv
\sum\limits_{C(x,t)}E_{C(x,t)}e^{-E_{C(x,t)}/T}/G(t) $ and $
\widehat{E}_{2}(x,t) \equiv
\sum\limits_{C(x,t)}E_{C(x,t)}^{2}e^{-E_{C(x,t)/T}}/G(t). $
It is clear that $ \langle E \rangle=\sum\limits_{x}\widehat{E}(x,t) $ and
$ \langle E^{2} \rangle=\sum\limits_{x}\widehat{E}_{2}(x,t). $
The derivation for the iteration relations for $ \widehat{E}(x,t) $ and $
\widehat{E}_{2}(x,t) $ are given as follows: \\
\vskip 0.02in
\noindent $ \widehat{E}(x,t+1)=\sum\limits_{C(x,t+1)}E_{C(x,t+1)} 
e^{-E_{C(x,t+1)/T}}/G(t+1) $ \\
\vskip 0.02in
\noindent $  =[\sum\limits_{C(x-1,t)}(E_{C(x-1,t)}+{\epsilon}_{l}) 
e^{-(E_{C(x-1,t)}+{\epsilon}_{l})/T}+\sum\limits_{C(x+1,t)}
(E_{C(x+1,t)}+{\epsilon}_{r})e^{-(E_{C(x+1,t)}+{\epsilon}_{r})/T}]/G(t+1)
$ \\
\vskip 0.02in
\noindent $
=[e^{-{\epsilon}_{l}/T}\widehat{E}(x-1,t)G(t)+e^{-{\epsilon}_{r}/T} 
\widehat{E}(x+1,t)G(t)+{\epsilon}_{l}e^{-{\epsilon}_{l}/T}G(x-1,t) 
+{\epsilon}_{r}e^{-{\epsilon}_{r}/T}G(x+1,t) ]/G(t+1) $ \hfill (A1) \\
\vskip 0.02in
\noindent and \\ 
\vskip 0.02in
\noindent $ \widehat{E}_{2}(x,t+1)=\sum\limits_{C(x,t+1)}E_{C(x,t+1)}^{2} 
e^{-E_{C(x,t+1)/T}}/G(t+1) $ \\
\vskip 0.02in
\noindent $  =[\sum\limits_{C(x-1,t)}(E_{C(x-1,t)}+{\epsilon}_{l})^{2} 
e^{-(E_{C(x-1,t)}+{\epsilon}_{l})/T}+\sum\limits_{C(x+1,t)}
(E_{C(x+1,t)}+{\epsilon}_{r})^{2}e^{-(E_{C(x+1,t)}+{\epsilon}_{r})/T}]/G(t+1)
$ \\
\vskip 0.02in
\noindent $
=[e^{-{\epsilon}_{l}/T}\widehat{E}_{2}(x-1,t)G(t)+e^{-{\epsilon}_{r}/T} 
\widehat{E}_{2}(x+1,t)G(t)+2{\epsilon}_{l}e^{-{\epsilon}_{l}/T}
\widehat{E}(x-1,t)G(t)+2{\epsilon}_{r}e^{-{\epsilon}_{r}/T}
\widehat{E}(x+1,t)G(t) $ \\
\vskip 0.02in
$  +{\epsilon}_{l}^{2}e^{-{\epsilon}_{l}/T}
G(x-1,t)+{\epsilon}_{r}^{2}e^{-{\epsilon}_{r}/T}G(x+1,t)]/G(t+1).
$ \hfill (A2) \\
\vskip 0.02in

\vskip 0.4in
\noindent {\bf References and Notes}
\vskip 0.2in
\begin{enumerate}

\item
Halpin-Healy, T.; Zhang, Y. -C.; {\sl Phys. Rep.} {\bf 1995}, 254, 215.
\item
Barab\'{a}si, A. -L.; Stanley, H. E.; {\sl Fractal Concepts in Surface
Growth}; Cambridge University Press: Cambridge, 1995.
\item
Meakin, P; {\sl Phys. Rep.} {\bf 1993}, 235, 189.
\item Kardar, M.; Parisi, G.; Zhang, Y. -C.; {\sl Phys. Rev. Lett.} {\bf
1996}, 56, 889. 
\item
Family, F.; Vicsek, T; {\sl Dynamics of Fractal Surfaces}; World
Scientific: Singapore, 1991.
\item
Huse, D. A.; Henly, C. L. {\sl Phys. Rev. Lett.} {\bf 1985}, 54, 2708.
Huse, D. A.; Henly, C. L. {\sl Phys. Rev. Lett.} {\bf 1985}, 55, 2924.
\item
Kardar, M.; {\sl Phys. Rev. Lett.} {\bf 1985}, 55, 2235.
\item
Kardar, M; Zhang, Y. -C.; {\sl Phys. Rev. Lett.} {\bf 1987}, 58, 2087.
\item 
Krug, J; Meakin, P; Halpin-Healy, T; {\sl Phys. Rev. A} {\bf 1992}, 45,
638.
\item
Kert\'{e}sz, J.; Horv\'{a}th V. K.; Weber F.; {\sl Fractals} {\bf 1992},
1, 67.
\item
Derrida, B.; Spohn, H.; {\sl J. Stat. Phys.} {\bf 1988}, 51, 817.
\item
Bolle, C. A.; Aksyuk, V.; Pardo, F.; Gammel, P. L.; Zeldov, E.; Bucher
E.; Boie, R.; Bishop, D. J.; and Nelson, D. R.; {\sl Nature} {\bf 1999},
399, 43-46.
\item
Schwartz, M;  Edwards, S. F.; {\sl Europhys. Lett.} {\bf 1992}, 20, 301.
Schwartz, M;  Edwards, S. F.; {\sl Phys. Rev. E} {\bf 1998}, 57, 5730.
\item
Moser, K.; Kert\'{e}sz, J.; Wolf, D. E.; {\sl Physica A} {\bf 1991},
178, 217.
\item
Medina, E.; Hwa, T.; Kardar, M.; Zhang, Y. -C.; {\sl Phys. Rev. A} {\bf
1989} 39, 3053.
\item
Kim, J.; Bray, A. J.; and Moore, M. A.; {\sl Phys. Rev. A} {\bf 1991},
44, R4782.
\item
Pellegrini Y. P.; Jullien, R.; {\sl Phys. Rev. Lett.} {\bf 1990}, 64,
1745.
\item
Forster, D; Nelson, D. R.; Stephen, M, J,; {\sl Phys. Rev. A} {\bf 1977}
16, 732.
\item
Kim, J. M.; Moore, M. A.; Bray A. J.; {\sl Phys. Rev. A} {\bf 1991}, 44,
2345.
\item
Schwartz, N; Nazaryev, A. L.; Havlin, S.; {\sl Phys. Rev. E} {\bf 1998},
58, 7642.
\item
Krug, J.; Halpin-Healy T.; {\sl J. Phys. A} {\bf 1998}, 31, 5939.
\item
Tang, L. -H.; Nattermann, T.; Forrest. B. M.; {\sl Phys. Rev. Lett.}
{\bf 1990}, 65,2422.
\item
Forrest B. M.; Tang, L. -H.; {\sl Phys. Rev. Lett.} {\bf 1990}, 64, 1405. 
\item
Derrida, B.; Golinelli, O; {\sl Phys. Rev. A} {\bf 1990}, 41, 4160.
\item
Press, W. H.;  Teukolsky, S. A.; {\sl Computers in Physics} {\bf 1992},
6, 522.
\item
Edwards, S. F.; and Wilkinson, D. R.; {\sl Proc. Roy. Soc. London A} {\bf
1982}, 381, 17.
\item
Honda, K; {\sl Fractals} {\bf 1996}, 4, 331.
\item
Honda, K.; {\sl Phys. Rev. E} {\bf 1997}, 55, R1235.
\item
Amar, J. G.; Family, F.; {\sl Phys. Rev. A} {\bf 1992}, 45, 5378.
\item
Fisher, D. S.; Huse, D. A.; {\sl Phys. Rev. B} {\bf 1991}, 43, 10728.
\item
Wang, X. H.; Havlin S.; Schwartz, M.; {\sl Entropy fluctuations in
directed polymers: indication for a phase transition in 2+1 dimensions},
preprint.
  
\end{enumerate}


\vfill\break
\onecolumn

\begin{figure}
\epsfig{file=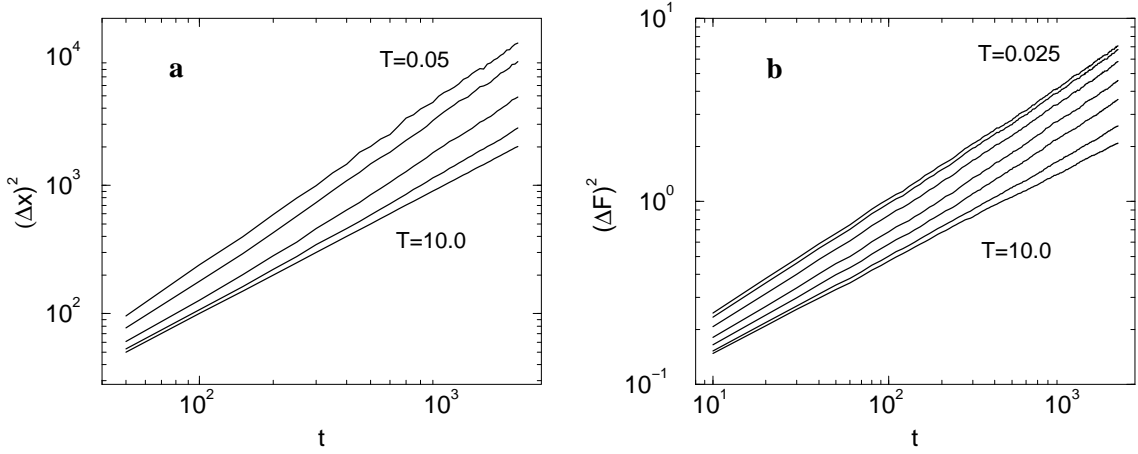,width=8cm,clip=,bbllx=0,bblly=0,bburx=350,
bbury=800,angle=-90}
\caption{ (a) Plot of the transverse fluctuations $ (\Delta x)^{2} $ as a
function of time $ t $ at different temperatures $ T=10, 1, {1\over 2},
{1\over 4}, {1\over 20} $ (from bottom to top) in $ d=1+1 $ dimensions.
(b) Plot of the free energy fluctuations $ (\Delta F)^{2} $ as a function
of time $ t $ at different temperatures $ T=10, 1, {1\over 2}, {1\over 3},
{1\over
5}, {1\over 10} {1\over 40} $ (from bottom to top) in $ d=1+1
$ dimensions.}
\end{figure}

\begin{figure}
\epsfig{file=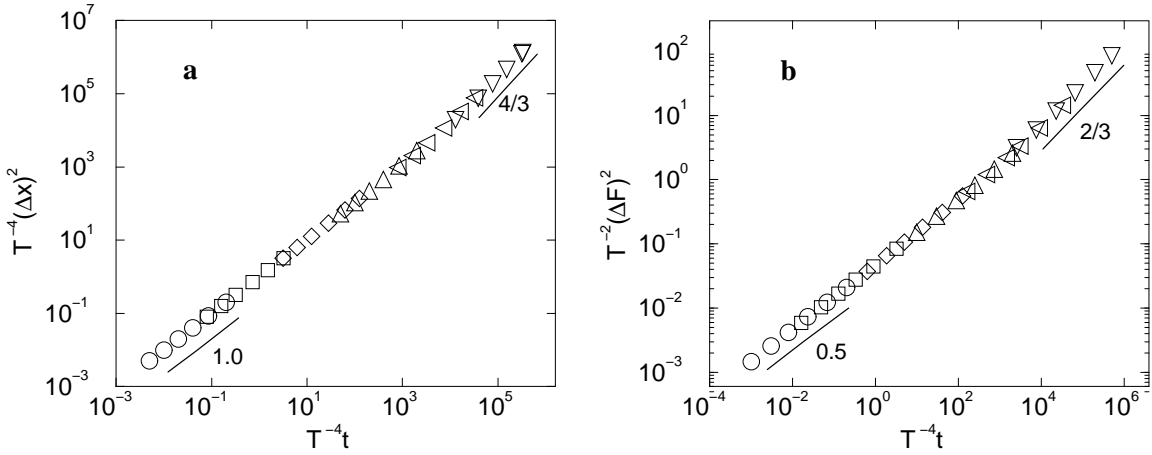,width=8cm,clip=,bbllx=0,bblly=0,bburx=350,
bbury=800,angle=-90}
\caption{ Scaling plot for the transverse fluctuations $ (\Delta x)^{2} $
shown in Fig. 1a and the free energy fluctuations $ (\Delta F)^{2} $ shown
in Fig. 1b for $ d=1+1 $ dimensions where circle, square, diamond,
triangle up, triangle left, triangle down correspond to different
temperatures $ T=10, 5, 2, 1.0, 0.5, 0.25 $
respectively. }
\end{figure}

\begin{figure}
\epsfig{file=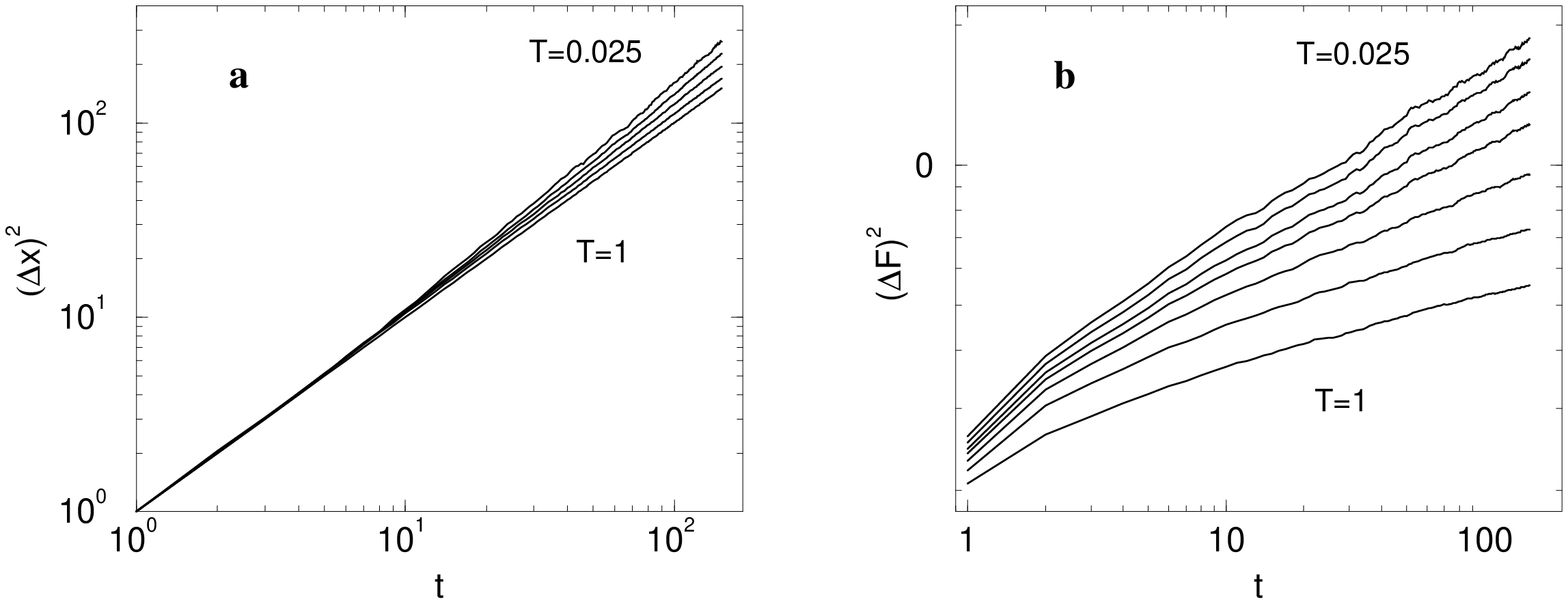,width=8cm,clip=,bbllx=0,bblly=0,bburx=350,
bbury=800,angle=-90}
\caption{ (a) Plot of the transverse fluctuations $ (\Delta x)^{2} $ as a
function of time $ t $ at different temperatures $ T=1, {1\over 5},
{1\over 7}, {1\over 10}, {1\over 40} $ (from bottom to top) in $ d=2+1 $
dimensions.
(b) Plot of the free energy fluctuations $ (\Delta F)^{2} $ as a function
of time $ t $ at different temperatures $ T=1, {1\over 4}, {1\over 6},
{1\over 8}, {1\over 10}, {1\over 15}, {1\over 40} $ (from bottom to top)
in $ d=2+1 $ dimensions.}
\end{figure}

\begin{figure}
\epsfig{file=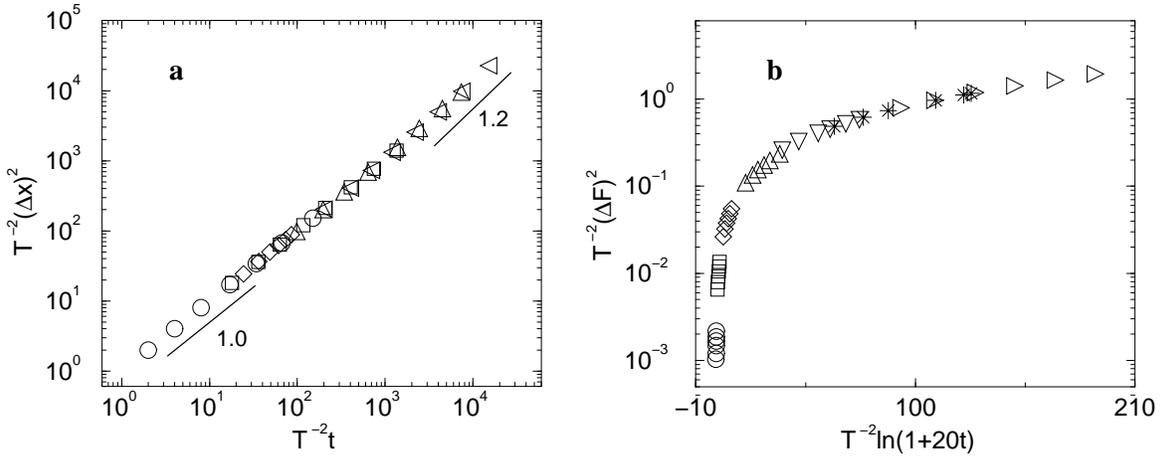,width=8cm,clip=,bbllx=0,bblly=0,bburx=350,
bbury=800,angle=-90}
\caption{ (a) Scaling plot for the transverse fluctuations 
$ (\Delta x)^{2} $ in $ d=2+1 $ dimension where circle, square, diamond,
triangle up, triangle left, correspond to different temperatures $ T=1,
{1\over 3}, {1\over 5}, {1\over 7}, {1\over 10} $ respectively. (b)
Scaling plot for the free energy fluctuations $ (\Delta F)^{2} $ in $
d=2+1 $ dimensions where circle, square, diamond, triangle up, triangle
down, star, triangle right correspond to different temperatures $ T=5, 1,
{1\over 2}, {1\over 3}, {1\over 4}, {1\over 5} $ respectively.}
\end{figure}

\begin{figure}
\epsfig{file=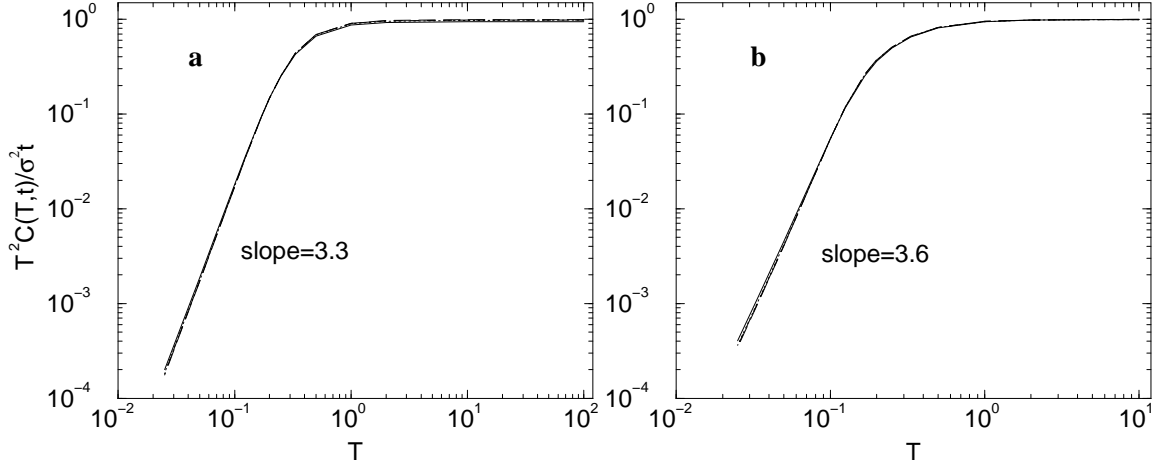,width=8cm,clip=,bbllx=0,bblly=0,bburx=350,
bbury=800,angle=-90}
\caption{ Plot of the specific heat as a function of temperature,
(a) for the different time $ t=100, 500, 1000, 2000 $ in $ d=1+1 $ 
dimensions and (b) for the different time $ t=50, 100, 120, 150 $ 
in $ d=2+1 $ dimensions. }
\end{figure}

\begin{figure}
\epsfig{file=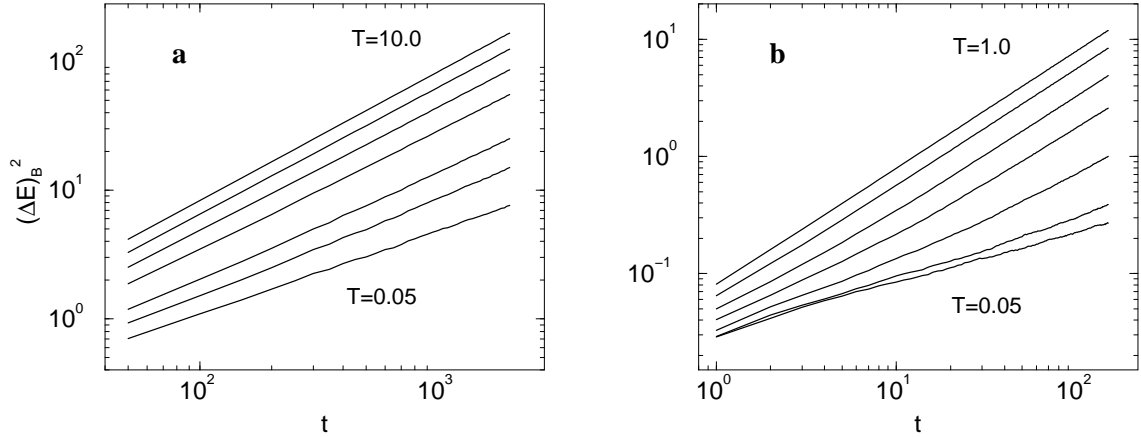,width=8cm,clip=,bbllx=0,bblly=0,bburx=350,
bbury=800,angle=-90}
\caption{ Plot of the ensemble energy fluctuations $ (\Delta E)^{2}_{B}
$ as a function of time $ t $, (a) for different temperatures $ T=10,
{1\over 2}, {1\over 3}, {1\over 4}, {1\over 6}, {1\over 8}, {1\over 20}  $
(from top to bottom) in $ d=1+1 $ dimensions and
(b) for different temperatures $ T=1, {1\over 3}, {1\over 5},
{1\over 7}, {1\over 10}, {1\over 15}, {1\over 20} $ (from top to bottom)
in $ d=2+1 $ dimensions.}
\end{figure}

\begin{figure}
\epsfig{file=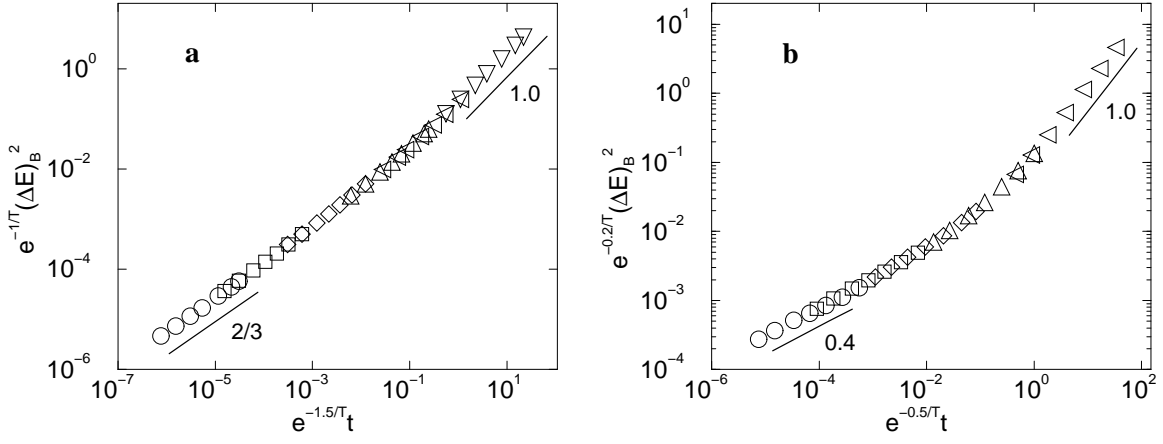,width=8cm,clip=,bbllx=0,bblly=0,bburx=350,
bbury=800,angle=-90}
\caption{ (a) Scaling plot for the ensemble energy fluctuations $
(\Delta E)^{2}_{B} $ in $ d=1+1 $ dimensions where circle, square,
diamond, triangle up, triangle left, triangle down correspond to different
temperatures $ T={1\over 12}, {1\over 10}, {1\over 8}, {1\over 6}, {1\over
5}, {1\over 3} $ respectively. 
(b) Scaling plot in $ d=2+1 $ dimensions where circle, square, diamond,
triangle up, triangle left correspond to $ T= {1\over 25}, {1\over 20},
{1\over 15}, {1\over 10}, {1\over 3} $ respectively. }
\end{figure}

\begin{figure}
\epsfig{file=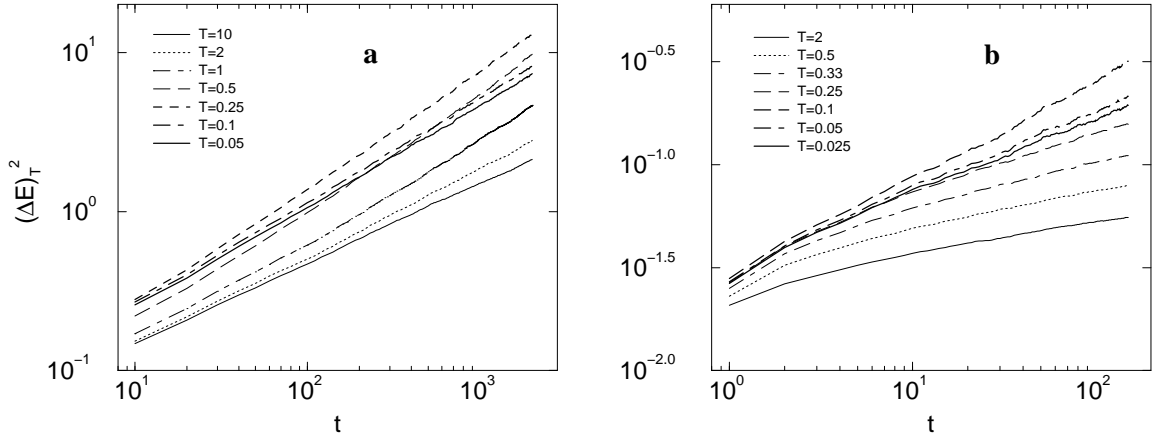,width=8cm,clip=,bbllx=0,bblly=0,bburx=350,
bbury=800,angle=-90}
\caption{ Plot of the internal energy fluctuations $ (\Delta E)^{2}_{T}
$ as a function of time $ t $, (a) in $ d=1+1 $ dimensions
and (b) in $ d=2+1 $ dimensions.}
\end{figure} 

\begin{figure}
\epsfig{file=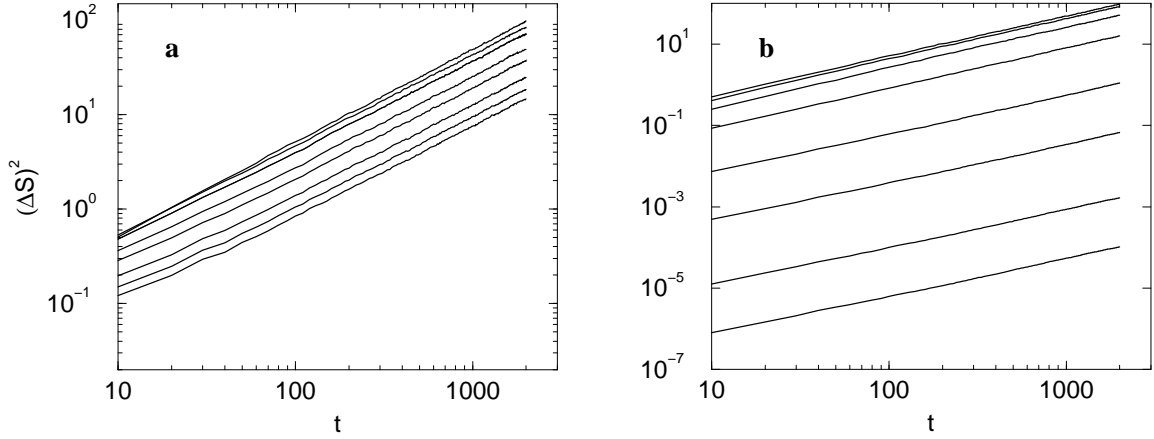,width=8cm,clip=,bbllx=0,bblly=0,bburx=350,
bbury=800,angle=-90}
\caption{ Plot of entropy fluctuations $ (\Delta S)^{2} $ as a function of
time $ t $ in $ d=1+1 $, (a) for different temperatures $ T={1\over 5},
{1\over 8}, {1\over 10}, {1\over 15}, {1\over 20}, {1\over 30}, {1\over
40}, {1\over 50} $ (from top to bottom) and (b) for different
temperatures $ T=10, 5, 2, 1, {1\over 2}, {1\over 3}, {1\over 4}, {1\over
5} $ (from bottom to top).}
\end{figure}

\begin{figure}
\epsfig{file=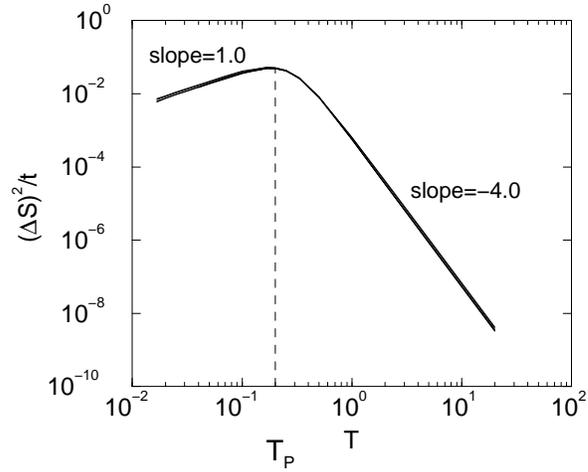,width=8cm,clip=,bbllx=0,bblly=0,bburx=350,
bbury=800,angle=-90}
\caption{ Plot of entropy fluctuations per unit length of the polymer
$ (\Delta S)^{2}/t $ as a function of temperature $ T $ for the different
time $ t=50, 500, 1000, 2000 $ in $ d=1+1 $ dimensions.}
\end{figure}

\begin{figure}
\epsfig{file=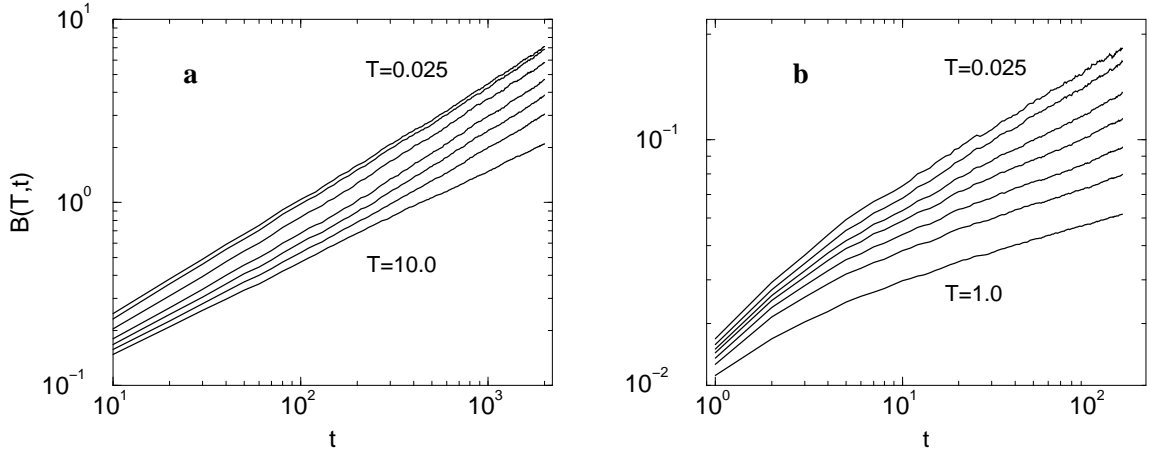,width=8cm,clip=,bbllx=0,bblly=0,bburx=350,
bbury=800,angle=-90}
\caption{ Plot of the function $ B(T,t) $ given in the equation (5) of
the text as a function of time $ t $, (a) for different temperatures $
T=10, 1, {2\over 3}, {1\over 2}, {1\over 3}, {1\over 5}, {1\over 40} $
(from bottom to top) in $ d=1+1 $ dimensions and (b) for different
temperatures $ T=1, {1\over 3}, {1\over 4}, {1\over
5}, {1\over 6}, {1\over 8}, {1\over 40} $ (from bottom to top) in $ d=2+1
$ dimensions.} 
\end{figure}

\end{document}